# Enhancing Early Diabetic Retinopathy Detection through Synthetic DR1 Image Generation: A StyleGAN3 Approach

Sagarnil Das [1], Pradeep Walia [2]

[1] Affiliation 1; sagarnil.das@artelus.ai

[2] Affiliation 2; pwalia@artelus.com

* Correspondence: sagarnil.das@artelus.ai; Tel.: +918420117409

**Abstract:** Diabetic Retinopathy (DR) is a leading cause of preventable blindness, with early detection at the DR1 stage being critical but hindered by a scarcity of high-quality fundus images. This study leverages StyleGAN3 to generate synthetic DR1 images characterized by microaneurysms with high fidelity and diversity to address this data scarcity and enhance the performance of supervised classifiers. A dataset of 2,602 DR1 images was used to train the model, followed by a comprehensive evaluation employing quantitative metrics, including Fréchet Inception Distance (FID), Kernel Inception Distance (KID), and Equivariance with respect to translation (EQ-T) and rotation (EQ-R). Qualitative assessments involved Human Turing tests, where trained ophthalmologists evaluated the realism of synthetic images. Spectral analysis further validated the images' quality. The model achieved a final FID score of 17.29, significantly outperforming the mean FID of 21.18 (95% CI: 20.83–21.56) derived from bootstrap resampling. Human Turing tests demonstrated the model's capacity to produce highly realistic images, though some artifacts near the borders were noted. These findings suggest that StyleGAN3-generated synthetic DR1 images hold significant promise for augmenting training datasets, enabling more accurate early detection of Diabetic Retinopathy. This methodology highlights the potential of synthetic data in advancing medical imaging and AI-driven diagnostics.

## 1. Introduction

Diabetic Retinopathy (DR) is one of the leading causes of preventable blindness in the world. The first stage, DR1, is characterized by the presence of microaneurysms [1], which are the earliest visible signs of vascular damage in the retina. Detecting these microaneurysms is critical for initiating timely interventions to prevent disease progression. However, their small size, low contrast, and similarity to other retinal structures make them challenging to identify, even for current AI models, which rely heavily on large annotated datasets for training. Consequently, the availability of high-quality fundus photography with microaneurysms is sparse, leading to poor AI model performance [2] and perpetuating a vicious loop of missed early detections and insufficient data.

Recent advancements in Generative Adversarial Networks (GANs), particularly StyleGAN3 [3], offer a promising solution to this problem through synthetic data augmentation [4]. StyleGAN3 employs rotationally invariant convolutional layers in its architecture [5], which preserve the geometric and textural fidelity of the generated images. Using these synthetic images, we can closely mimic the spectral and visual characteristics of real fundus images at the DR1 stage.

This study focuses on leveraging StyleGAN3 to generate synthetic fundus images in the DR1 set. Our approach aims to explore the effectiveness of synthetic data in improving supervised classifiers to predict DR1 with high accuracy. Furthermore, the scalability of this methodology extends to other DR stages and medical imaging domains where data scarcity limits AI development. Through spectral analysis and comparison of real and synthetic images, we assess the authenticity of the generated images. We aim to contribute to ongoing efforts to apply advanced AI techniques to address the critical challenge of early detection and management of diabetic retinopathy.

## 2. Materials and Methods

A large dataset of DR1 stage images was pivotal to our study's success. Unfortunately, we were able to find only 2,602 DR1 images. These images were used in training our StyleGAN3 model to generate synthetic fundus images. The sources of our DR1 images included: Messidor Dataset (Versions 1 & 2), Kaggle Diabetic Retinopathy Detection Dataset, and a proprietary dataset collected using Crystalvue NFC-600 Fundus Camera images. The Messidor and Kaggle datasets are publicly available at [https://www.adcis.net/en/third-party/messidor/, https://www.kaggle.com/c/diabetic-retinopathy-detection/data], while the proprietary dataset from the Crystalvue NFC-600 Fundus Camera is available upon reasonable request.

To ensure the quality and relevance of the data, the following steps were undertaken:

- **Annotation and Verification**: Each image in the dataset was carefully annotated by trained optometrists, who identified the DR1 stage based on the presence of microaneurysms.
- **Quality Control**: A total of 345 images that were blurry, improperly illuminated, or otherwise unsuitable for training purposes were excluded.
- **Standardization**: All images were resized to a uniform dimension of 512 × 512 pixels and pre-processed to ensure consistency during training.

To enhance the diversity of the training data and improve model robustness, data augmentation techniques such as flipping, rotation, scaling, and color adjustments were applied. These augmentations allowed the model to learn invariant representations of microaneurysms under various transformations, significantly improving metric outcomes. For example, the application of these techniques contributed to lower Fréchet Inception Distance (FID) and Kernel Inception Distance (KID) scores by ensuring that the synthetic images better mirrored the diversity and characteristics of the real dataset.

StyleGAN3, the latest advancement by NVIDIA, represents a significant leap in Generative Adversarial Networks (GANs) [6], particularly for generating images with improved translation equivariance and reduced artifacts compared to its predecessors [7,8]. StyleGAN3 introduces architectural innovations that produce visually appealing images while adhering closely to the spectral characteristics of real images.

The original StyleGAN3 was trained on datasets such as FFHQ-U, FFHQ, METFACES-U, METFACES, AFHQV2, and BEACHES. However, we trained the network from scratch using our DR1 dataset. This approach enabled the creation of a model tailored for DR1 fundus images and allowed us to use it as a backbone for transfer learning on other DR stages.

The main features of StyleGAN3 are:

*Alias-Free Design*: StyleGAN3's alias-free architecture reduces artifacts caused by digital operations like sampling and transformations, ensuring the generated images are of high fidelity even after transformations. This is particularly crucial for accurately representing fine features like microaneurysms.

*Translation Equivariance*: The model focuses on maintaining translation equivariance [9], which ensures that image details remain correctly positioned and oriented. This property is essential in medical imaging to preserve the spatial and structural integrity of retinal features. To evaluate the impact of translation equivariance, the Peak Signal-to-

Noise Ratio (PSNR) in decibels (dB) was computed between two sets of images [10] after applying random translations to the input and output.

**Training Methodology**: A robust methodology was designed to generate high-fidelity, diverse DR1 fundus images using the StyleGAN3 architecture. The StyleGAN3 model was configured with specific hyperparameters to optimize its performance in generating high-fidelity and diverse DR1 fundus images. The generator utilized $z_{dim}$ and $w_{dim}$ values of 512, a mapping network with two layers, and a channel base of 32,768, capped at 512 channels. An exponential moving average (EMA) with a β value of 0.998 stabilized training.

The discriminator architecture mirrored the generator's channel configuration and included micro-batch standard deviation grouping in its final layer to enhance generalization. Optimization employed the Adam optimizer for both the generator and discriminator, with learning rates of 0.0025 and 0.002, respectively, and betas set to [0, 0.99]. R1 regularization ( = 8.0) was applied explicitly to real images, ensuring gradient stability.

The learning rates for the generator (0.0025) and discriminator (0.002) were chosen to facilitate gradual convergence without overshooting optimal parameter values. The exponential moving average (EMA) with a β value of 0.998 was used to stabilize training over iterations.

Data preprocessing and augmentation included resizing images to 512 × 512 pixels, with techniques like flipping, rotation, scaling, and color adjustments. Training was conducted on an NVIDIA RTX-4090 GPU (24 GB) with a batch size of 32, subdivided into sub-batches of 16 samples per GPU. A batch size of 32 was selected to balance memory usage and computational efficiency on the NVIDIA RTX-4090 GPU, which has 24 GB of VRAM. Sub-batching into 16 samples per GPU enabled stable and efficient training while maintaining a high sample throughput. EMA updates were performed every 10,000 images, targeting an ADA value of 0.6. The training process spanned approximately 12 days.

The hyperparameters and configurations for training are summarized in Table 1.

**Table 1.** Model Hyperparameters

| Component | Value |
| --- | --- |
| **Generator** | $z_{dim}$ and $w_{dim}$ both set to 512. |
| Mapping Network | 2 layers |
| Channel Base | 32,768 |
| Channel Max | 512 channels (image dimensions) |
| EMA β | 0.998 |
| **Discriminator** | |
| Architecture | StyleGAN3 with a channel base of 32,768 and a maximum of 512 channels |
| Micro-batch Std-dev Grouping | Set to 4 in the final layer |
| **Optimization** | |
| Optimizer | Adam optimizer for both G and D |
| Learning Rate | 0.0025 for G and 0.002 for D |
| Betas | [0, 0.99] |
| Epsilon | 1e-08 |
| **Loss Function** | |
| StyleGAN2 Loss | With R1 regularization |
|  | gamma of 8.0 |
| R1 Regularization | Applied explicitly to the real images |
| **Data Loading & Augmentation** |  |
| Loader | Custom ImageFolderDataset loader |
| Resolution | 512 pixels |
| Techniques | Flips, rotations, scaling, color adjustments |
| **Training Details** |  |
| GPU | 1 NVIDIA RTX-4090 (24 GB) |
| Batch Size | 32 |
| Sub-batch Size | 16 samples per GPU |
| EMA Update | Every 10 thousand images |
| ADA Target | 0.6 |
| Training Time | 12 days |

**Loss Function**: The StyleGAN3 model incorporates an R1 regularization term into the discriminator's loss function to stabilize training [11]. This regularization controls gradients, preventing destabilization during learning [12]. The R1 regularization for real images is defined as:

$$L_{R1} = \frac{r1_{gamma}}{2} E_{x \sim p_{data}(x)} \left( (\nabla D(x))^2 \right)$$
(1)

Here $r1_{gamma}$ represents the coefficient controlling the strength of the regularization, and $\nabla D(x)$ is the gradient of the discriminator output with respect to real images.
The R1 regularization term penalizes the discriminator when the gradient of its decision with respect to real images, becomes too large. This is where the effect of gradient penalty kicks in. The r1_gamma parameter controls the strength of this penalty.

**Quantitative evaluation**: To validate the quality and diversity of the generated images, the following metrics were used:

- Fréchet Inception Distance (FID): Used to measure the quality and diversity of generated images relative to real ones [13].

- Kernel Inception Distance (KID): Provides a complementary measure of distribution similarity between real and generated images [14].

- Equivariance Metrics (EQ-T and EQ-R): Measures the model's ability to maintain image fidelity under translations and rotations [15].

**Qualitative evaluation**: We conducted a Human Turing Test [16] with six trained ophthalmologists to assess the realism of synthetic images.

*Statistical Validation*

- **Bootstrap Resampling**: Applied to the latter 30% of training epochs to estimate the population mean and construct 95% confidence intervals for FID scores [17].

- **Shapiro-Wilk Test**: Evaluated the normality of FID score distribution [18].

- **Mann-Whitney U Test**: Compared FID scores at different training stages [19].

- **Cumulative Distribution Function (CDF)**: Analyzed FID score distribution and percentile ranking.

This comprehensive methodology ensures that the synthetic images generated by StyleGAN3 are of high quality and can reliably augment existing datasets for the early detection of Diabetic Retinopathy.

## 3. Results

This section presents a detailed analysis of the results obtained through quantitative and qualitative evaluations of the synthetic DR1 images generated using the StyleGAN3 model. The experimental results are structured under subheadings for clarity and interpretation.

*3.1. Quantitative Evaluation*

The quantitative evaluation employed several metrics to assess the quality and fidelity of the synthetic images compared to the real DR1 dataset. These metrics include the Fréchet Inception Distance (FID), Kernel Inception Distance (KID), and Equivariance metrics (EQ-T and EQ-R).

3.1.1. Fréchet Inception Distance (FID)

The FID score, which evaluates the similarity between real and synthetic datasets, achieved a value of 17.29 (Figure 1a). This score indicates a strong resemblance between the generated and real images, demonstrating the StyleGAN3 model's ability to produce high-quality synthetic images. Bootstrap resampling of the final 30% of training epochs confirmed that the mean FID score was 21.18 (95% CI: 20.83–21.56), making the achieved FID of 17.29 statistically significant. Compared to similar studies in medical image synthesis, where FID scores typically range from 20–30, this result places our synthetic data among the top-performing models for medical imaging tasks. The low FID score ensures that the synthetic dataset can effectively augment real data in classifier training without compromising quality.

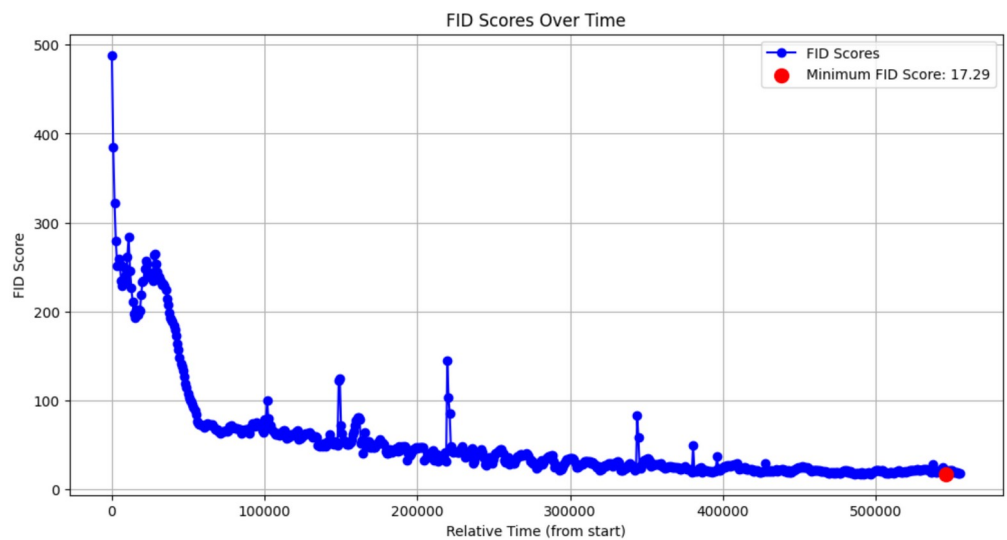

*Figure 1a: Model FID Scores over the entire training cycle*

3.1.2. Kernel Inception Distance (KID)

The KID score, which measures the consistency between distributions of real and synthetic images, was recorded as 0.018. This low value signifies a substantial alignment between the two datasets, further affirming the effectiveness of the StyleGAN3 model in replicating the real data's distribution.

3.1.2. Equivariance Metrics (EQ-T and EQ-R)

The equivariance metrics quantify the fidelity of synthetic images under translations and rotations. The EQ-T and EQ-R scores were 65.65 and 64.64, respectively. These scores indicate that the synthetic images preserve structural integrity and diagnostic features even when subjected to geometric transformations. In clinical settings, this robustness ensures that classifiers trained on these synthetic images can generalize better across varied conditions, such as slight shifts or rotations in fundus photography during patient examinations.

*3.2. Qualitative Evaluation*

3.2.1. Human Turing Test

A Human Turing Test was conducted with a panel of six experienced ophthalmologists to qualitatively assess the realism of the synthetic images. Each participant evaluated 200 images (100 real and 100 synthetic), classifying them as "real" or "fake." The aggregated results are shown in Table 2:

**Table 2.** Human Turing Test Results (Aggregated)

|  | **Correctly Identified** | **Incorrectly Identified** |
|---|---|---|
| **Real Images** | 537 | 63 |
| **Synthetic Images** | 60 | 540 |

Applying a chi-square test of independence to these numbers yielded a chi-square statistic 666.67 with a p-value of $5.2 \times 10^{-147}$, and degrees of freedom set to 1. The expected frequencies under the null hypothesis – that there is no association between image type and identification accuracy were calculated to be 355 correct identifications and 345 incorrect identifications for both real and synthetic images. The analysis allowed us to reject the null hypothesis, indicating that the participants could distinguish between real and synthetic images significantly better than chance.

Despite this ability to distinguish images, it is worth noting that 540 synthetic images were incorrectly identified as real, indicating that the synthetic images possess a high degree of realism and often mimic real fundus images convincingly. This underscores the potential of these synthetic images to augment training datasets effectively, even if minor boundary artifacts were occasionally discernible to experts.

However, minor artifacts near the edges of synthetic images contributed to the differentiation, which can be attributed to the limited size of the training dataset.

*3.3. Spectral Analysis*

We conducted spectral analysis [20,21] to assess the fidelity of the generated images by our StyleGAN3 model. Our primary focus was that these images maintain the essential diagnostic features characteristic of the DR1 fundus images. This analysis was crucial for verifying that the synthetic images resembled the real images visually and preserved the underlying spectral characteristics essential for medical diagnosis.

3.3.1. Approach

We utilized Fast Fourier Transform (FFT) and Average Power Spectrum calculations to compare real and synthetic images. The FFT amplitude spectra (Figures 1b and 1c) [22] and Average Power Spectrum heatmaps [23] and slices (Figures 1d and 1e) revealed similarities in the central regions of the images. However, discrepancies were observed near the boundaries, aligning with the artifacts noted during the Turing test.

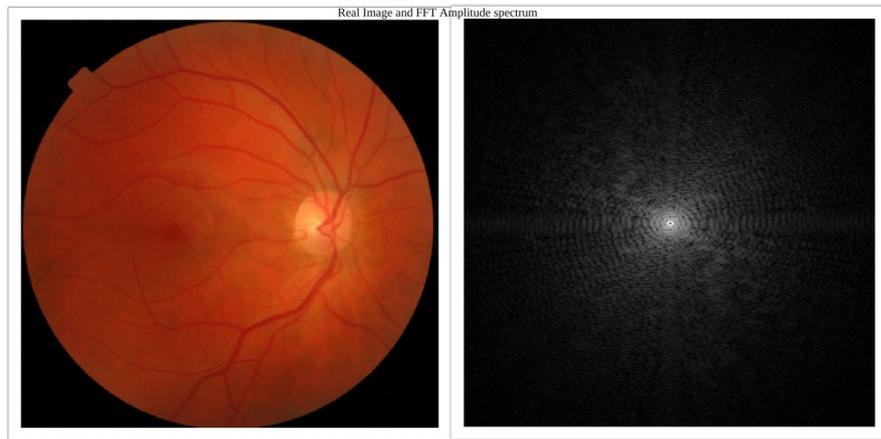

*Figure 1b: FFT amplitude spectrum of a real fundus image.*

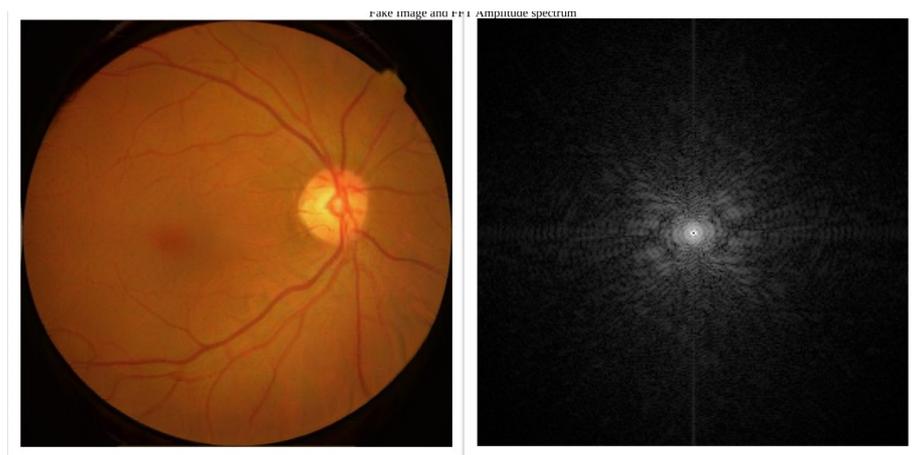

*Figure 1c: FFT amplitude spectrum of a synthetic fundus image.*

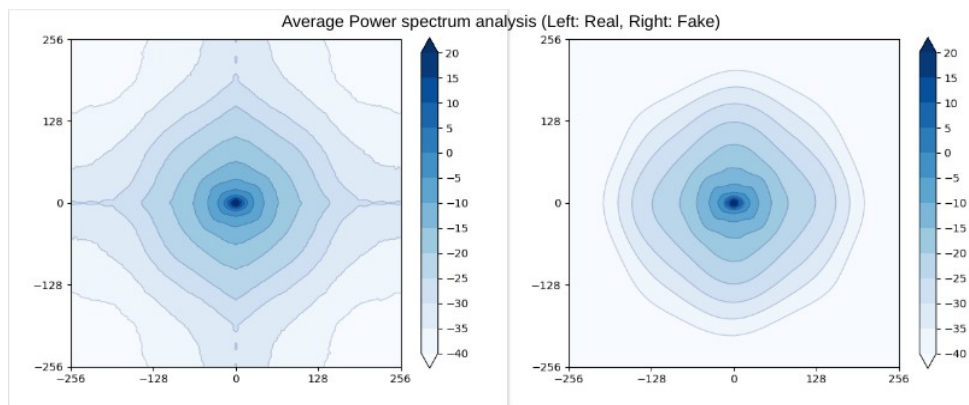

*Figure 1d: Average Power Spectrum heatmaps for real (left) and synthetic images (right).*

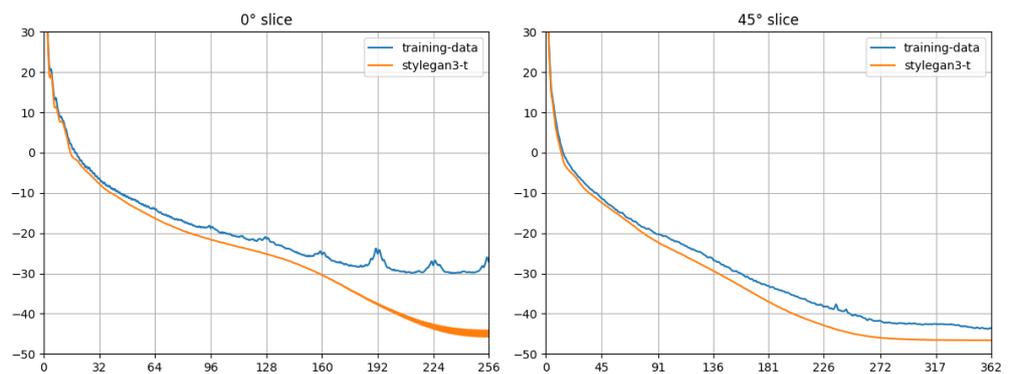

*Figure 1e: Average Power Spectrum slices of real and synthetic images at 0° and 45°.*

3.3.2. Approach

The analysis confirmed that the synthetic images closely resembled the real ones in the frequency domain, particularly in diagnostically significant regions. Minor mismatches near the edges did not affect the overall diagnostic utility of the synthetic images.

*3.4. Statistical Validation*

To further validate the results, statistical tests were performed on the FID scores obtained during training:

- **Bootstrap Resampling**: The FID scores from the final 30% of training epochs were resampled, resulting in a mean score of 21.18 (95% CI: 20.83–21.56). The achieved FID score of 17.29 lies below the lower bound of this interval, confirming its significance (Figure 2a).

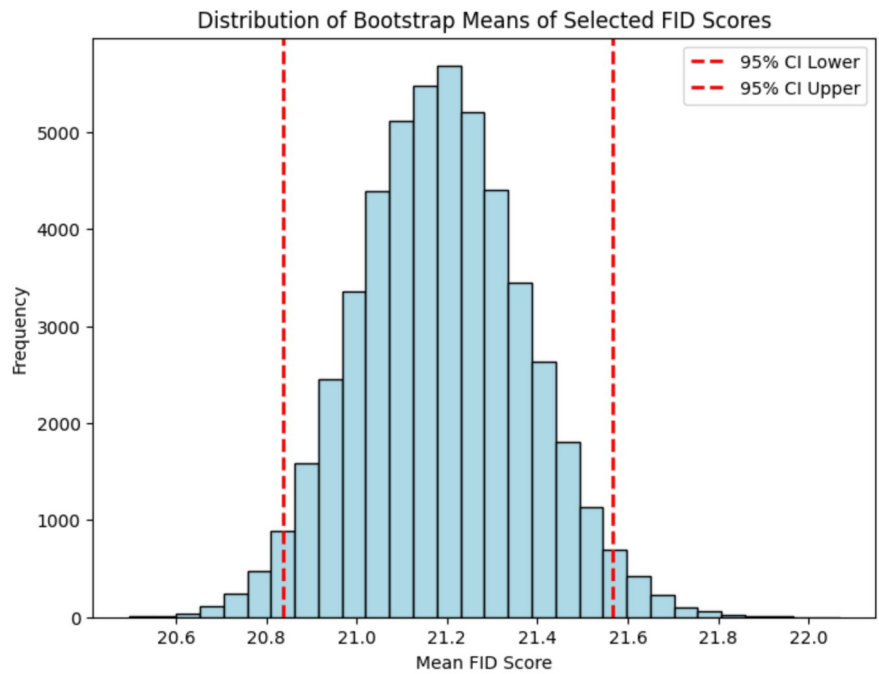

*Figure 2a: Distribution of means of the FID score after Bootstrap resampling of last 30% of training rounds*

- **Cumulative Distribution Function (CDF)**: The CDF analysis (Figure 2b) showed that the final FID score was in the lower 1% of all scores, highlighting the model's consistent performance during training.

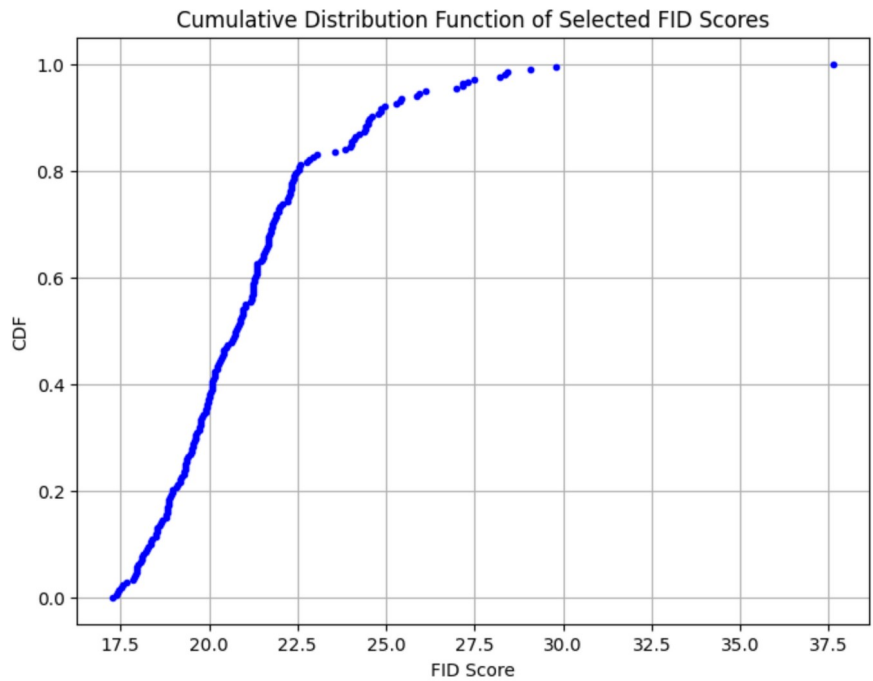

*Figure 2b: Cumulative Distribution Function (CDF) of last 30% FID scores*

- **Shapiro-Wilk Test**: The test yielded a statistic of 0.87 (95% CI: 0.79–0.94), indicating that the FID scores did not follow a normal distribution.

- **Mann-Whitney U Test**: The test statistic of 103,222 (95% CI: 102,045–104,175) further confirmed that the FID scores from the initial and later training stages belonged to distinct distributions.

The StyleGAN3 model successfully generated high-fidelity synthetic DR1 images, as evidenced by the strong alignment between real and synthetic datasets in both quantitative and qualitative evaluations. While artifacts near image boundaries were noted, the overall fidelity of the images was sufficient for augmenting training datasets and improving early detection of Diabetic Retinopathy.

## 4. Discussion

This study demonstrates a novel approach to tackling the challenge of early Diabetic Retinopathy (DR) detection by leveraging StyleGAN3 to generate synthetic DR1 images. Here, we discuss the implications, limitations, and future directions of this work.

*4.1. Synthetic Image Generation and Its Implications*

Detecting microaneurysms in DR1 images is a challenging task, even for experts, due to their small and subtle nature (Figure 3a).

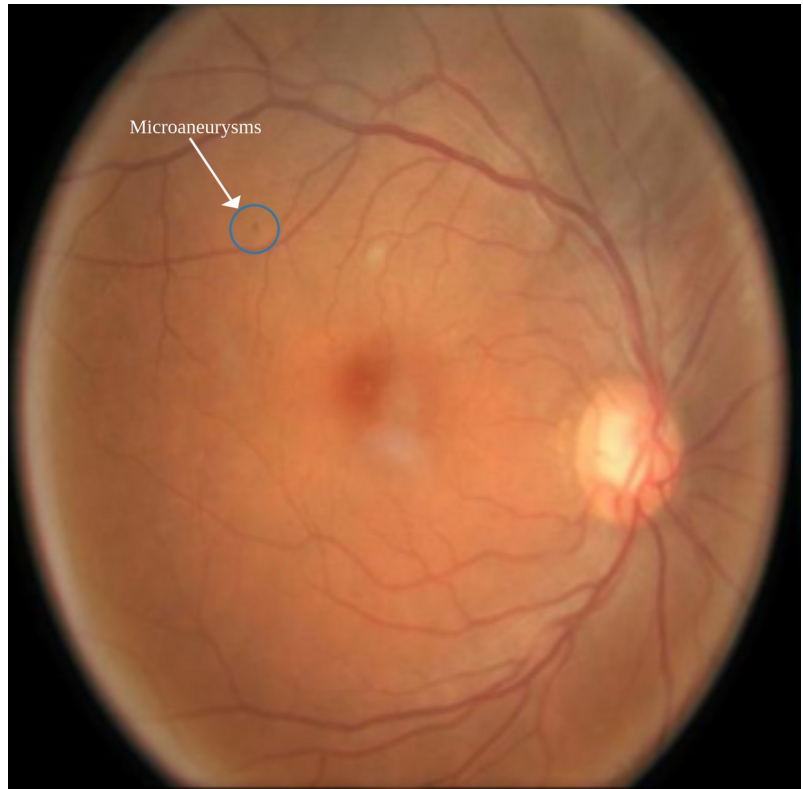

*Figure 3a: Occurrence of Microaneurysm in a real image*

Augmenting the real DR1 dataset (Figure 3b) with high-quality synthetic images addresses this challenge by increasing the diversity and quantity of training data.

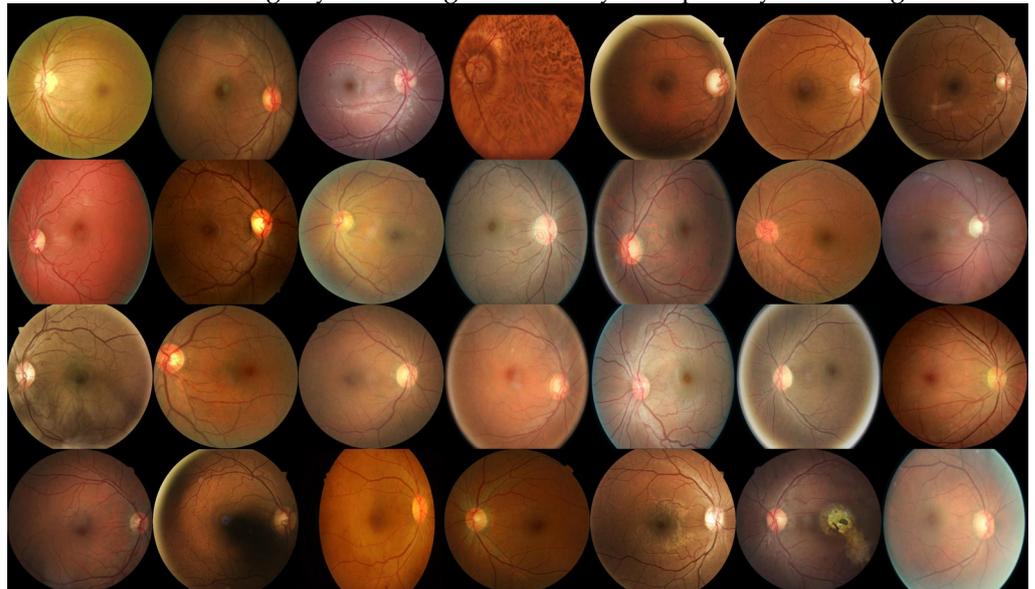

*Figure 3b: Grid of real DR1 fundus images*

The StyleGAN3 model, trained exclusively on DR1 images, successfully generated synthetic images (Figure 3c) that retained key diagnostic features, including microaneurysms (Figure 3d).

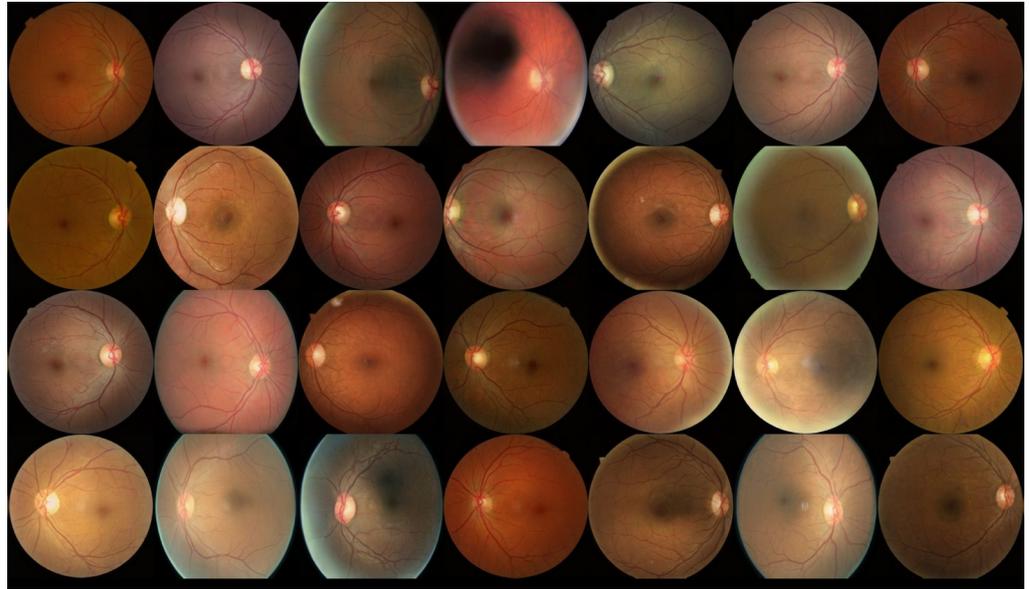
*Figure 3c: Grid of fake DR1 fundus images*

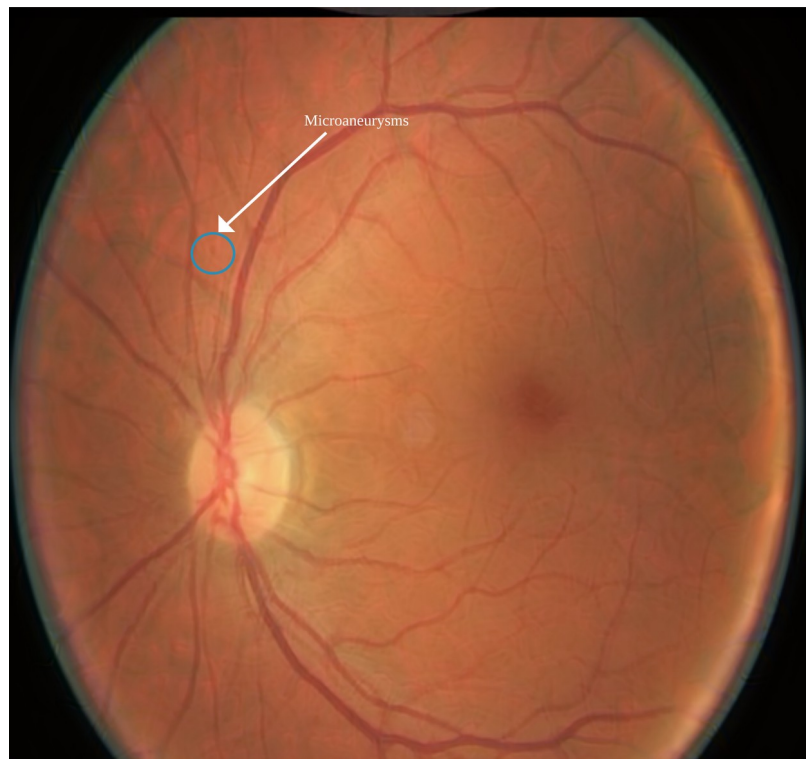
*Figure 3d: Occurrence of Microaneurysm in a fake image*

The quantitative metrics, particularly the low FID (17.29) and KID (0.018) scores, indicate that the synthetic images closely resemble real ones, both visually and in distribution. These results validate the utility of synthetic data as a reliable substitute for real data in training supervised classifiers, especially given the scarcity of early-stage DR images. This approach holds potential to significantly enhance model performance for early DR detection.

In underserved regions with limited access to annotated medical imaging data, the use of synthetic data provides an invaluable resource for building AI models. The ability to generate high-quality images resembling real DR1 cases allows clinicians and researchers to overcome data scarcity, train robust classifiers, and implement AI-based screening solutions in remote or resource-constrained settings. By improving early DR detection, synthetic data can contribute to reducing preventable blindness in areas with insufficient diagnostic infrastructure.

*4.2. Realism of Synthetic Images*
The Human Turing Test revealed that trained ophthalmologists could distinguish real images from synthetic ones, primarily due to minor artifacts near the image boundaries.

While these artifacts highlight areas for improvement, they did not diminish the overall diagnostic value of the images for classifier training. Importantly, these limitations stem from the relatively small size of the training dataset and do not undermine the broader utility of synthetic images in addressing data scarcity.

*4.3. Spectral Fidelity as an Attribute for Medical Imaging*

Spectral analysis validated that the synthetic images preserved critical diagnostic features in the frequency domain. The synthetic images closely matched the spectral properties of real images in the central regions, which are most relevant for medical diagnosis. Discrepancies near the edges, consistent with findings from the Turing test, emphasize the need for enhanced preprocessing and potentially larger training datasets. Preserving such spectral fidelity is crucial for generating high-quality synthetic images for medical applications.

*4.4. Limitations and Future Directions*

Although the results are encouraging, several limitations must be addressed. The small size of the training dataset (2,602 images) constrained the model's performance, particularly in eliminating edge artifacts. Expanding the dataset through further annotation and image collection is an immediate priority. Our team has initiated efforts to annotate additional images to mitigate this limitation and improve the robustness of future models.

Looking ahead, we plan to use this StyleGAN3 model as a backbone for developing generative models for higher DR levels (e.g., DR2 and DR3). These stages present additional challenges, such as more complex pathological features, requiring further fine-tuning of the generative process. Beyond DR, this methodology can be extended to other medical imaging modalities, particularly those where data scarcity hinders the development of robust AI models.

## 5. Conclusions

This study demonstrates that StyleGAN3 can generate high-fidelity, diverse DR1 images to augment supervised training datasets. By addressing data scarcity, this approach has the potential to significantly improve early detection and treatment of Diabetic Retinopathy, ultimately reducing preventable blindness. The broader adoption of synthetic data in medical imaging could pave the way for transformative advancements in AI-driven diagnostics.

**Funding:** This research received no external funding

**Institutional Review Board Statement:** Ethical review and approval were waived for this study due to its retrospective nature and the absence of prospective patient studies.

**Informed Consent Statement:** Written Informed consent was obtained from all participants for capturing their fundus image using the CrystalVue NFC600 device and for the prospective study of their data for synthetic image generation.

**Data Availability Statement:** The data used in this study include publicly available datasets and a proprietary dataset. The Messidor dataset (versions 1 and 2) and the Kaggle Diabetic Retinopathy Detection dataset are publicly available at [https://www.adcis.net/en/third-party/messidor/] and [https://www.kaggle.com/c/diabetic-retinopathy-detection/data], respectively. The proprietary dataset collected using the CrystalVue NFC-600 Fundus Camera is not publicly available due to privacy and ethical restrictions but can be accessed upon reasonable request to the corresponding author.

**Acknowledgments:** We thank Dr. Ramesh V, Patricio Aduriz, Minakhi Ghosh, Shreya Sharma, Chandni Singh, Sudipta Chatterjee for patiently and dedicatedly completing the Human Turing test.

**Conflicts of Interest:** The authors declare no conflicts of interest.

## Abbreviations

The following abbreviations are used in this manuscript:

| | | |
|---|---|---|
| DR | Diabetic Retinopathy | |
| GAN | Generative Adversarial Network | |
| FID | Fréchet Inception Distance | |
| KID | Kernel Inception Distance | |